\documentclass[final,3p,times,twocolumn]{elsarticle}
\usepackage{graphicx} %provides the includegraphics command
\usepackage{amssymb} %provides various useful mathematical symbols
\usepackage{color}
\usepackage{lineno} %adds line numbers; start line numbering with \begin{linenumbers}, end it with \end{linenumbers} or switch it on for the whole article with \linenumbers after \end{frontmatter}.
\usepackage{hyperref}
\usepackage{nicefrac}

\setcitestyle{authoryear,open={(},close={)}} %Citation-related commands
\biboptions{semicolon,round,sort}
\journal{}
\begin{document}
%%%%%%%%%%%%%%%%%%%%%%%%%%%%%%%%%%%%%%%%%%%%%%%%%%%%%%%%%%%%%
\begin{frontmatter}
%%%%%%%%%%%%%%%%%%%%%%%%%%%%%%%%%%%%%%%%%%%%%%%%%%%%%%%%%%%%%
\title{Restoring the structure: A modular analysis of\\ ego-driven organizational networks}
% \title{Modular analysis of ego-driven organizational networks} 

\author{Robert P. Dalka\fnref{label1,label2}}
\ead{rpdalka@umd.edu}
\author{Justyna P. Zwolak\corref{cor1}\fnref{label2}}
\ead{jpzwolak@nist.gov}
\cortext[cor1]{Corresponding author.}
\address[label1]{Department of Physics, University of Maryland, 
College Park, MD 20742, USA}
\address[label2]{National Institute of Standards and Technology, 
%Applied and Computational Mathematics Division, 
Gaithersburg, MD 20899, USA}
%%%%%%%%%%%%%%%%%%%%%%%%%%%%%%%%%%%%%%%%%%%%%%%%%%%%%%%%%%%%%
%% use the tnoteref command within \title for footnotes;
%% use the tnotetext command for the associated footnote;
%% use the fnref command within \author or \address for footnotes;
%% use the fntext command for the associated footnote;
%% use the corref command within \author for corresponding author footnotes;
%% use the cortext command for the associated footnote;
%% use the ead command for the email address,
%% and the form \ead[url] for the home page:
%%
%% \title{Title\tnoteref{label1}}
%% \tnotetext[label1]{}
%% \author{Name\corref{cor1}\fnref{label2}}
%% \ead{email address}
%% \ead[url]{home page}
%% \fntext[label2]{}
%% \cortext[cor1]{}
%% \address{Address\fnref{label3}}
%% \fntext[label3]{}

%%%%%%%%%%%%%%%%%%%%%%%%%%%%%%%%%%%%%%%%%%%%%%%%%%%%%%%%%%%%%
\begin{abstract}
Organizational network analysis (ONA) is a method for studying interactions within formal organizations.
The utility of ONA has grown substantially over the years as means to analyze the relationships developed within and between teams, departments, and other organizational units.
The mapping and quantifying of these relationships have been shown to provide insight into the exchange of information and resources, the building of social capital, and the spread of culture within and between organizations.
However, the ethical concerns regarding personally identifiable information (PII) that exist for traditional social science research are made more pertinent in ONA, as the relational nature of the network may leave participants open to identification by organization management.
To address this, we propose a method of generating a network of organizational groups (e.g. units, departments, teams) through the projection of ego-networks absent of PII.
We validate this method through modular analysis of the resulting networks and compare the identified structure to a known structure of the organization.
The methodology lays a foundation for performing ONA that needs only anonymous ego-centric data to identify large-scale aspects of organizational structures. 
\end{abstract}
%%%%%%%%%%%%%%%%%%%%%%%%%%%%%%%%%%%%%%%%%%%%%%%%%%%%%%%%%%%%%
%\begin{keyword}
%Science \sep Publication \sep Complicated
%% keywords here, in the form: keyword \sep keyword

%% MSC codes here, in the form: \MSC code \sep code
%% or \MSC[2008] code \sep code (2000 is the default)
%\end{keyword}

\end{frontmatter}
%%%%%%%%%%%%%%%%%%%%%%%%%%%%%%%%%%%%%%%%%%%%%%%%%%%%%%%%%%%%%
% \linenumbers %start line numbering here if you want
%%%%%%%%%%%%%%%%%%%%%%%%%%%%%%%%%%%%%%%%%%%%%%%%%%%%%%%%%%%%%
\section{Introduction}
%%%%%%%%%%%%%%%%%%%%%%%%%%%%%%%%%%%%%%%%%%%%
Society is built around social interactions, driven by individuals desire to form and maintain meaningful relationships \citep{Tchalova15-HBF}.
Social interactions -- be it personal, family- or work-related -- help people build communication skills and cooperate to achieve common goals.
From a methodological standpoint, the social structures resulting from such interactions can be investigated using social network analysis (SNA) \citep{Wasserman94,Scott11}.
In the context of formal organizations, the application of SNA is called organizational network analysis (ONA). 
ONA is an empirical research method for mapping, analyzing, and quantifying relationships used to perform work between individuals, groups, and whole organizations \citep{Merrill07-ONA}.
The key feature distinguishing ONA from more traditional survey-based data analytic methods is the use of structural or relational variables and analysis techniques based on graph-theoretic methods. 
ONA allows the creation of statistical and graphical models of the people, tasks, and groups forming the network.
It also allows to assess and quantify how the interactions between them are facilitated by complex structural connections to distribute knowledge and resources important to organizational systems \citep{Kilduff10-OSN}. 

ONA can provide insights into social influences within teams, the effect of team building on the dynamics of an organization’s social network, or identify cultural issues within an organization \citep{Blau17, Cook78-PEC, Krackhardt90-APL, Bavelas50-CPG, Quintane13-SLS, Flap98-ION}.
An analysis of formal and informal relationships in organizations can, in turn, help shape business or research strategies by maximizing the organic exchange of information, thereby helping the organization become more effective and innovative \citep{Gulati02-ONB, Cross04, Colombo11-OIF, Ozman09-IFN}.
Detecting communities in networks can reveal how individual nodes form groups that work together to perform larger functions to fulfill an organization's mission, e.g., intra- and inter-unit collaborations, shared professional experiences, similar working functions, as well as capture network structures within and between organizations \citep{Brandes09-NES, Lazega12-NSD, Carpenter12-SNO}.
The utility of ONA to analyze and understand the various relationships developed within and between teams, departments, and other organizational units has grown substantially over the years \citep{Moliterno11-NTM}.
To date, ONA has been used to investigate individual employees networks to understand structural social capital, access to resources, and the spread of attitudes and culture \citep{Borgatti03-NPO}; to understand the relationship between multiple levels of networks, such as the individual and their group, division, or organization affiliations \citep{Moliterno11-NTM, Brass04-SNO, Lazer03-BEN}; and to investigate the network structure between organizations \citep{Provan07-INL}.

With the growing prevalence of ONA, there has been increased interest in the ethical issues associated with social sciences research and the investigation of personal and organizational networks \citep{Agneessens22-SCS, Borgatti05-TEG, Ellison17-SMR, Morris15, Cronin21-EIN}.
The codes of ethical conduct typically focus on preventing harm to participants, guaranteeing privacy and anonymity, providing informed consent, and avoiding deception \citep{Diener78, Nijhawan13-IIC, Harris08-CAC}.
While anonymity is the most powerful device for protecting the interest of respondents, ensuring confidentiality of individual responses in ONA can be challenging \citep{Bell07-EMR}.
While the standard practice of anonymizing data prior to analysis helps to address this issue, it does not completely eliminate it. 
For example, minority employees can often be uniquely identified given even a small number of attributes.
Moreover, the relational nature of network data means that the survey data often includes information about third parties (the so-called {\it alters}) who, depending on the sampling strategy, might not have given their consent or even been informed of their indirect participation in the study \citep{Borgatti03-ESI}. 
At the same time, constructing a full organizational network necessitates an inclusion of not only the respondents but also the alters and interactions between them as perceived by the respondent.
Thus, while ONA can help leaders understand whether there are sufficient connections to facilitate positive relationships needed for innovative teams, the requirement to provide personally identifiable information (PII) necessary to establish the complete ONA makes this approach challenging to implement in certain settings (e.g., within government institutions).

To overcome this challenge, we demonstrate an alternative method for assessing the broadly defined interconnectedness within an organization. 
In particular, we show that a network built by projecting anonymous egocentric network data (ego-networks) onto organizational units provides a valid and reliable approximation of a known organizational structure.
We use ties between employees to indicate ties between their respective working units as an example of the duality between individuals and groups \citep{Brass04-SNO, Breiger74-DPG}.
We then perform a modularity analysis of the resulting network and confirm that egocentric data representing individuals and their connections, but containing no PII, is sufficient to approximate the complete network structure of organizational units.

We test the utility of the proposed approach using the ``National Institute of Standards and Technology (NIST) Inclusivity Network'' dataset pertaining to two types of direct interaction reported by NIST's employees \citep{Espinal21-NIR}.
By design, the dataset does not contain PII about the ego or the alters and the specific operating units affiliations are anonymized prior to analysis.
To build the complete network of NIST's organizational units we employ an affiliation projection method.  
The structure of the resulting network, established through common community detection techniques, is validated against the known structure of the organization.
We also investigate the stability of the identified communities through two data re-sampling experiments to further verify the effectiveness of our approach.
The ability to use anonymous ego-networks to generate inter-unit networks broadens the applicability of ONA, particularly to institutions where privacy concerns make analysis using PII impossible.

This paper is organized as follows: In Section~\ref{sec:methods}, we introduce our methodology in broad terms, including the creation of the network, the community detection analysis and the evaluation of those communities.
The results specific to the data collected through the NIST Inclusivity Network survey, as well as the two experiments, are presented in Section~\ref{sec:results}.
Section~\ref{sec:discussion} presents a discussion of the results and the potential limitations of the proposed approach.
Finally, we conclude with a discussion in Section~\ref{sec:conclusions}.

%%%%%%%%%%%%%%%%%%%%%%%%%%%%%%%%%%%%%%%%%%%%%%%%%%%%%%%%%%%%%%
\section{Methodology}\label{sec:methods}
%%%%%%%%%%%%%%%%%%%%%%%%%%%%%%%%%%%%%%%%%%%%
Network analysis uses {\it nodes} (e.g., individuals, families, departments at a university, or operational units in a company) and {\it ties} (e.g., friendship, kinship, co-authoring papers, or collaborating on a project) to represent social phenomena as a network.
Depending on the type of interactions defining the network, ties can have directionality and weight.
The directionality indicates whether or not a given interaction is mutual (e.g., kinship, co-authoring papers) or perceived as reciprocated (e.g., friendship).
Weights are used to quantify the value associated with each tie, providing additional information on the nature of the interaction (e.g., the frequency, strength, or cost of a given interaction).

The focus of our study is determining to what extent anonymous ego-network data can be used as a proxy to capture the organizational network structure. 
In this context, the nodes represent units in an organization and the ties represent interpersonal interactions reported by the employees (egos).

%%%%%%%%%%%%%%%%%%%%%%%%%%%%%%%%%%%%%%%%%%%%
\subsection{Creating organizational networks}
\label{ssec:ego-proj}
\begin{figure}[t]
\centering
\includegraphics{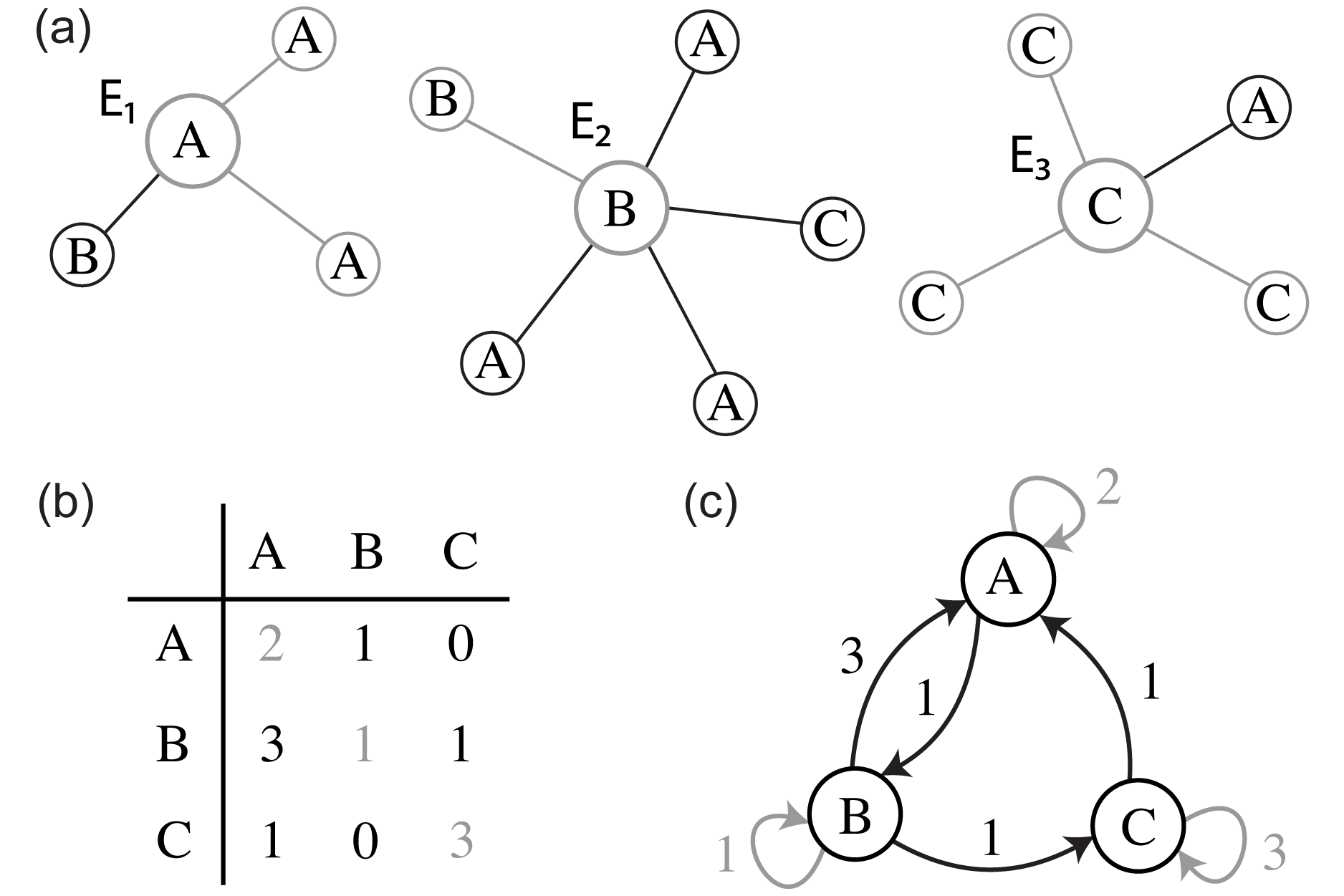}
\caption{Visualization of the process of projecting ego networks onto organizational units.
(a) Three sample ego networks with nodes, indicating respondents (big nodes) and alters (smaller nodes), labeled based on the unit affiliation.
(b) The incidence matrix built from networks shown in (a).
(c) The resulting weighted, directed organizational network.
The gray nodes, numbers, and arrows indicate the intra-unit interactions while the black color denotes inter-unit connections.}
\label{fig:net-cartoons}
\end{figure}

%%%%%%%%%%%%%%%%%%%%%%%%%%%%%%%%%%%
The social network data used in our analysis, described in more detail in Section~\ref{ssec:survey}, consists of two ego-network datasets each representing a distinct type of interaction self-reported by employees: one related to achieving work-related goals and a second related to seeking advice about career-related decisions \citep{Espinal21-NIR}.
Additional ego and alter characteristics, such as certain demographics, work history, and affiliation are included in the data.
However, unlike typical egocentric data, our datasets do not account for the relationships among the alters as perceived by the ego.
Figure~\ref{fig:net-cartoons}(a) shows a cartoon representation of three ego-networks (${\rm E}_1$, ${\rm E}_2$, and ${\rm E}_3$) like the ones included in our dataset, with nodes labeled based on a hypothetical unit affiliation A, B, or C. 

As further explained in Section~\ref{ssec:survey}, the data from the survey is stored as a response incidence matrix with rows representing respondents and columns indicating the number of alters from each unit a given respondent listed on their survey.
To generate the organizational networks, we aggregate the incidence matrix by egos' unit.
The rows of the resulting matrix represent the cumulative number of alters from each unit that egos from a given unit reported collaborating with.
Since the total number of interactions reported from unit A to unit B can be different than the total number of interactions from B to A, the resulting incidence matrix is asymmetric, indicating a weighted and directed network.
An incidence matrix built from the three networks shown in Fig.~\ref{fig:net-cartoons}(a) is depicted in Fig.~\ref{fig:net-cartoons}(b) and the resulting projected network is shown in Fig.~\ref{fig:net-cartoons}(c).

%%%%%%%%%%%%%%%%%%%%%%%%%%%%%%%%%%%
%%%%%%%%%%%%%%%%%%%%%%%%%%%%%%%%%%%

%%%%%%%%%%%%%%%%%%%%%%%%%%%%%%%%%%%%%%%%%%%%
\subsection{Network modularity}\label{ssec:sna-comm}
%%%%%%%%%%%%%%%%%%%%%%%%%%%%%%%%%%%
Detecting communities in networks can provide important insights about how individual nodes form groups that work together to perform a larger function. 
These communities can be defined as groups of nodes that have many strong ties between them and few weak ties to other groups.
The measurement of how well the network is divided into communities is called the modularity.
In our analysis, we have used the weighted undirected definition of modularity proposed by \cite{Newman04-AWN}.

In the creation of the undirected networks, we use the sum of ties in order to preserve information from all connections that are present in the directed networks.
This preserves connections that are not reciprocated in the undirected networks ensuring that connections originating from egos within small units are not lost, as these connections are the most likely to be not reciprocated.
In general, the function used to create the undirected network should be guided by the research objective.

The modularity for a partition of a weighted undirected network is given by,
\begin{equation}
Q = \frac{1}{2m}\sum_{i,j} \bigg[w_{ij} - \frac{C_{S,i} C_{S,j}}{2m}\bigg]\,\delta(c_i,c_j),
\end{equation}
where $w_{ij}$ represents the weight of a tie between nodes $i$ and $j$, $C_{S,i (j)}$ represents the strength of node $i$ ($j$), $c_{i (j)}$ represents the community to which node $i$ ($j$) belongs, and $m = \frac{1}{2}\sum_{i,j}w_{i,j}$ \citep{Newman04-AWN}.
The delta function, $\delta(c_i,c_j)$, equals 1 when $c_i = c_j$ and 0 otherwise.
The modularity ranges from -1 to 1 and compares the relative density of ties within communities to those ties between communities.
A positive value indicates a partition in which the ties within communities are more prevalent than those between communities.

The modularity can be used to determine how successful a particular partition is in dividing a network into communities.
It can also be used as a value to optimize for while creating the partition, as is the case in the Louvain community detection algorithm \citep{Blondel08-FUC}.
The Louvain algorithm consists of two stages.
The first stage starts by placing each node into its own community, then iteratively moves individual nodes into neighboring communities through choosing the merge that provides the greatest gain in modularity.
This process is repeated until there is no further increase to the modularity through reassignment of individual nodes.
The second stage in the algorithm then treats the communities detected in the first stage as nodes and the steps of the first stage are repeated.
Both stages are run iteratively until no further increase to modularity is possible.

We chose this community detection method because it creates communities that are maximally dense relative to ties between communities through a hierarchical model of community structure while accounting for the weight of ties between nodes.
There are many other widely used community detection algorithms, including those based on the removal of ties \citep{Newman04-FEC}, label propagation \citep{Raghavan07-NLT}, and random walks \citep{Rosvall08-MRW}.
In general, the choice of community detection algorithm should be matched to the specific characteristics of the network.

%%%%%%%%%%%%%%%%%%%%%%%%%%%%%%%%%%%%%%%%%%%%
\subsection{Validating community structures}
\label{ssec:exp}
%%%%%%%%%%%%%%%%%%%%%%%%%%%%%%%%%%%
In general, the response rate to a survey may vary between units, resulting in a skewed or biased dataset that is not representative of the organization.
In order to better understand how this may influence the communities that are constructed, we take various stratified samples from the full ego data.
The number of samples taken should be guided by the size of the dataset.
We ran experiments with samples from $N = 100$ to $N = 10\,000$ and found our results to converge at $N = 1\,000$.

To validate the community assignment, we run the Louvain community detection algorithm on each sampled network and use the resulting communities to build a {\it community membership matrix} $\bf{M}$. 
Entries of the matrix $\bf{M}$ represent shared community assignments for pairs of units, aggregated across all samples,
\begin{equation}
    M_{ij} = \frac{1}{N}\sum_{m = 1}^{N}\delta(L_{m,i}, L_{m,j}),
\end{equation}
with, $i,j=1,\dots,n$, where $n$ is the number of units, $N$ is the number of samples, and ${\bf L}=\{L_{m,i}\}_{N \times n}$ is a matrix in which the rows correspond to the community assignments for each sampled network and the columns correspond to a single unit.
The element values in $\bf{M}$ range between $0$, indicating that two units were never placed in the same community, and $1$, indicating that two units shared a community in each sample. 

Using the community membership matrix, we then quantify the {\it community assignment stability}, $S_{c}$, for each unit.
The score $S_{c}$ is defined as the percent difference between the average frequency of being partitioned with units from the expected community and the average frequency of being partitioned with units outside of that community.
$S_{c}=1$ when a given unit is always assigned to the expected community and decreases as the frequency of assignment to unexpected community increases.
At $S_{c}=0$, the chances of being assigned within and outside the expected community are equal.
At $S_{c}=0.5$, the chance of being assigned within the expected community is twice that of the unexpected community, which we define as {\it stable}.

Along with calculating $S_{c}$, we employ the community matching approach as described by \cite{Ghawi22-CMS} to calculate the average {\it purity} and {\it F-measure} of the samples as compared to the original community structure.
Purity and F-measure are extrinsic clustering evaluation metrics.
Purity quantifies the extent to which a community from partition A contains units from only a single community in the partition B.
The F-measure is made up of two quantities, {\it precision} and {\it recall}; precision of a community is the same as its purity, and recall captures the fraction of units that come from the community in partition A out of the total number of units in the community of partition B.
Each metric ranges from 0, indicating no overlap in partitions, to 1, indicating a perfect matching.

In our analysis, we define the expected community structure as the partitioning of the full network.
When available, the expected community structure should be defined as the true hierarchical structure of an organization. 

In addition to validating the community structure, we employ the community assignment stability score to determine the effect of time-dependent variations in response rates between units.
These variations may depend on the different workflows of each unit or group of units, resulting in disproportionate representations in the survey data.
To investigate these effects on the identified community structure, we perform an experiment treating our data as representing a complete organization.

In the experiment, we mimic distributed survey collections by sampling the majority of units at a rate typical of survey responses.
In addition, at each stage, a randomly selected subset of units is sampled at a lower rate to simulate the systematic differences in respondents' availability due to work-related constraints (e.g., important deadlines, particularly demanding phase of a project).
In practice, the frequency of sampling needs to be aligned with the expected organization's work schedule.

In our experiment, we chose to sample at four separate stages to simulate quarterly administration of the survey.
For each sampled network, we repeat the community assignment stability analysis and build the community membership matrices.
To determine at what point the communities become stable, the data is also analyzed cumulatively at each stage.

%%%%%%%%%%%%%%%%%%%%%%%%%%%%%%%%%%%%%%%%%%%%
\subsection{Test case: NIST Interactions Survey data}\label{ssec:survey}
%%%%%%%%%%%%%%%%%%%%%%%%%%%%%%%%%%%
The networks used in this study are built based on the NIST Interactions Survey data \citep{Espinal21-NIR}.
NIST is a non-regulatory agency of the United States Department of Commerce with a mission to ``promote innovation and industrial competitiveness by advancing measurement science, standards, and technology in ways that enhance economic security and improve quality of life'' \citep{NIST-info}.
The survey data was collected as part of one of three initiatives funded at NIST with the aim to better understand equity and inclusivity within the NIST workforce.
More information about the survey design as well as the results of the ego network analysis can be found in a NIST Interagency/Internal Report 8375 \citep{Espinal21-NIR}.

The NIST Interactions Survey data includes two datasets representing the two survey questions: one related to broadly defined collaborations across NIST and the other targeting the advice-related interactions.
Both datasets contain certain information about the respondents (e.g., gender, ethnicity, age group, tenure status, and the organizational unit to which they belong) as well as information about the alters listed by each ego (including their gender, age group relative to the ego, and their organizational unit).
We stress that no personally identifiable information was collected as part of the survey.
Thus, a direct mapping of the ego network data onto the full organization network, which would then enable analysis of the connectedness within the entire organization, is not directly possible.

The focus of this work is on recovering the organizational structure from anonymous ego network data. 
To build the network, we use the organizational unit information from egos and alters.
To prepare the dataset for our analysis, we first remove all of the egos' and alters' attributes reported in the survey except the unit affiliations and, in the case of egos, the unique ID.
We then create the incidence matrix by assigning to each ego the cumulative number of alters from each of the units. 
Since we are not interested in the specific nature of the individual units but rather their overall functions, the unit names are replaced with generic codes prior to the analysis.

NIST consists of 19 distinct organizational units, with six categorized as Laboratory Programs (coded as L1 through L6), three as Innovation and Industry Services (coded as P1 through P3), seven as Management Resources (coded as R1 through R7), and three serving more administrative functions (coded as M1 through M3) \citep{NIST-info}.
Since different units serve different functions within the organization, we expect that a network capturing the interconnectedness between the units should reveal a certain structure, with units of the same type being more connected than units of different types.

%%%%%%%%%%%%%%%%%%%%%%%%%%%%%%%%%%%%%%%%%%%%
\subsection{Visualization and Statistical Analysis}
\label{ssec:viz-stat}
%%%%%%%%%%%%%%%%%%%%%%%%%%%%%%%%%%%
All network visualizations and analyses presented in this work are carried out using the \textit{igraph} \citep{igraph} package in \textit{R} \citep{R}.
Visualizations of the community membership matrices is processed through the $R$ function, $\it{heatmap.2}$ \citep{gplots}.
The chi-square test is used to verify differences between the observed and expected percentages of intra-ties within units.
We consider results with $p < 0.001$ as significant.

%%%%%%%%%%%%%%%%%%%%%%%%%%%%%%%%%%%
\begin{figure*}[!ht]
\centering
\includegraphics{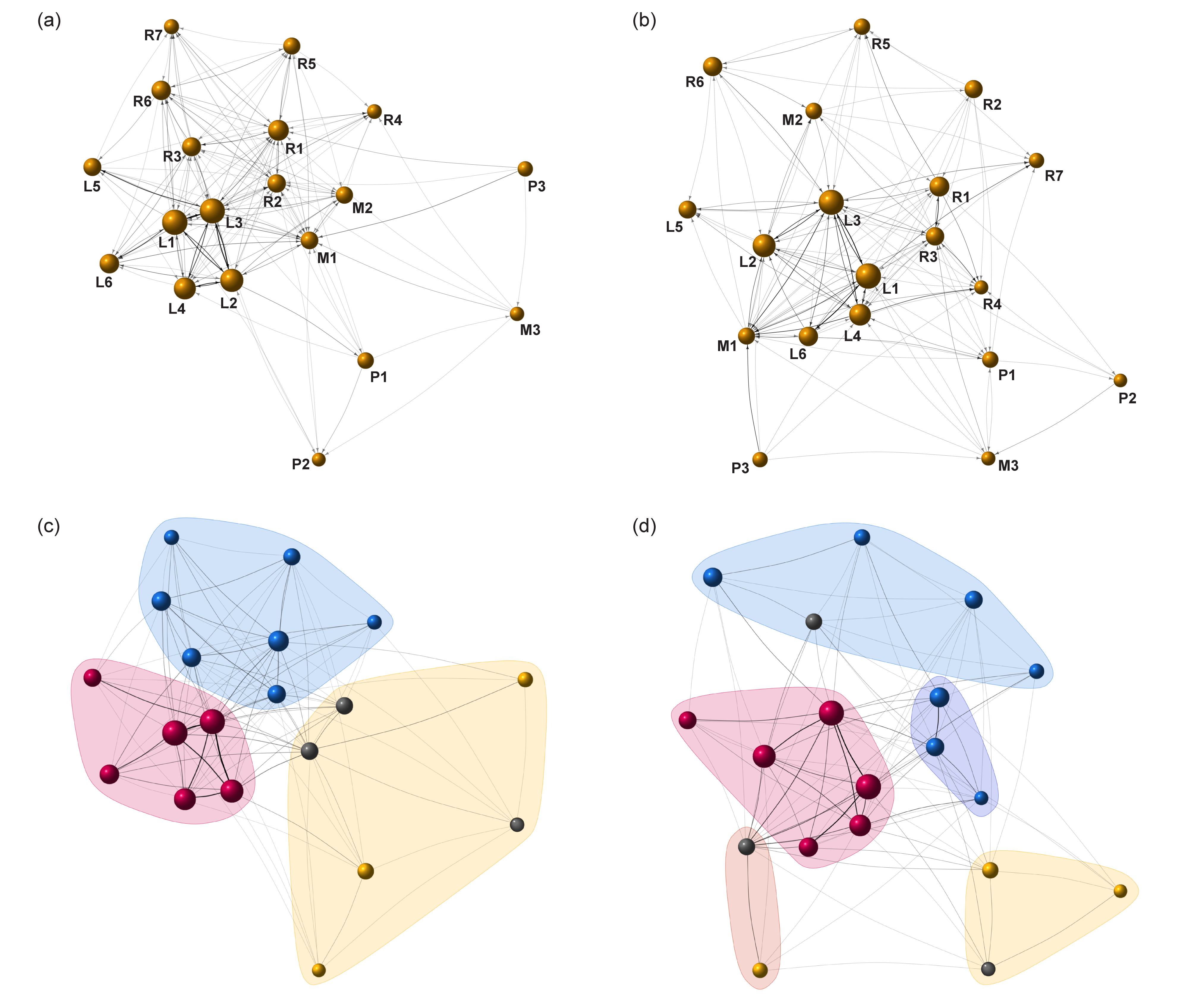}
\caption{Visualization of the weighted and directed network created based on (a) collaboration and (b) advice data and of the community structure for the (c) collaboration and (d) advice networks. 
In each plot, the nodes are sized by the number of responses from each unit and the line width corresponds to the weight of each tie.  
In the directed network, each unit is labeled. In the community plot, the nodes are colored by the organizational division to which that the particular unit belongs. 
The shaded region indicates the communities that are identified.}
\label{fig:networks-plots}
\end{figure*}
%%%%%%%%%%%%%%%%%%%%%%%%%%%%%%%%%%%

%%%%%%%%%%%%%%%%%%%%%%%%%%%%%%%%%%%%%%%%%%%%%%%%%%%%%%%%%%%%%%
\section{Results}\label{sec:results}
%%%%%%%%%%%%%%%%%%%%%%%%%%%%%%%%%%%%%%%%%%%%
The leading question of our study considers the efficacy of reproducing an entire organizational structure from anonymous ego-network data.
We perform a modularity analysis to identify the community structure within the resulting networks and test them against the known structure of the organization to validate the utility of the proposed approach.
We consider the extent to which our approach is successful by comparing the results of two types of network data gathered from the same population.

These two networks capture different aspects of connections between employees.
Whereas collaborating on a project necessitates interaction with other employees, not everyone seeks advice internally.
Moreover, while ``collaboration to achieve work goals'' should have a well defined and shared meaning among employees -- i.e., working together to fulfill the organization's mission -- seeking ``advice about important career-related decisions'' may be interpreted differently by various respondents.
Thus, we expect that the two resulting networks will have somewhat different organizational structures.

Finally, we investigate the stability of the community structures through a series of experiments aimed at understanding how the response rate effects the community structure and how the structure is effected over cumulative surveys.

%%%%%%%%%%%%%%%%%%%%%%%%%%%%%%%%%%%%%%%%%%%%
\subsection{Creating organizational networks}
%%%%%%%%%%%%%%%%%%%%%%%%%%%%%%%%%%%
As discussed in Section~\ref{ssec:survey}, the data from the survey is stored as a response incidence matrix with rows representing respondents and columns indicating the number of alters from each unit a given respondent listed on their survey.
To generate a weighted, directed network of the organizational units, we project the egocentric collaboration and advice networks onto units, following the process discussed in Section~\ref{ssec:ego-proj}.

To generate the two organizational networks, we aggregate the incidence matrix by egos' unit, separately for each dataset.
The resulting incidence matrices represent the weighted and directed collaboration and advice networks of units, as visualized in Fig.~\ref{fig:networks-plots}(a) and Fig.~\ref{fig:networks-plots}(b), respectively.

The organizational networks obtained through this process account for both intra-unit (i.e., between egos from a single unit) and inter-unit (i.e., between egos from different units) ties.
The frequency of intra-unit collaborations varies between units and between the types of networks.
In the collaboration network, on average $60.8~\%$ of egos (standard deviation ${\rm st.dev.}= 12.6~\%$) reported exclusively intra-unit ties, with the exact percentages ranging from $33.3\ \%$ (P3, $N=9$ respondents) to $76.3\ \%$ (L6, $N=38$ respondents).
For advice network, the fraction of egos reporting exclusively intra-unit ties was slightly higher, at $66.1~\%$ (${\rm st.dev.}= 18.4~\%$), with the per-unit percentages ranging from $33.3\ \%$ (M3, $N=6$) to $100\ \%$ (R7, $N=8$).
Table~\ref{tab:intra-stats} shows the number of respondents and the intra-connectedness per unit for both networks.
We find no correlation between the number of respondents from a given unit and the fraction of intra-unit ties for both networks ($\chi^2(18)=4.8$, $p=1$ for the collaboration network; $\chi^2(18)=8.0$, $p=0.98$ for the advice network).

Since due to the anonymity concerns, our data does not include information about more fine-grain affiliation (e.g., a specific division within the unit), we focus our analysis on the inter-unit ties. 
We note that more specific affiliation questions would enable mapping of the intra-unit subnetworks.
However, that might necessitate elimination of some of the demographic questions asked in the survey in order to preserve anonymity.
In our case, each organizational network represents the whole organization structure captured through the inter-unit interactions. 
The next step is to validate whether the obtained networks can serve as a proxy for the organizational structure.

%%%%%%%%%%%%%%%%%%%%%%%%%%%%%%%%%%%
\begin{table}[t]
\caption{The comparison of the number of egos from each unit who reported collaboration- and advice-related interactions, the percentage of egos who only reported intra-units ties $I_{in}$, and the community stability measure $S_{c}$.}
\centering
\begin{tabular}{l r r r r r r }
\hline
\textbf{Unit} & \multicolumn{3}{l}{\bf Collaboration} & \multicolumn{3 }{l}{\bf Advice} \\
 & Size & $I_{in}$ [\%] & $S_{c}$ & Size & $I_{in}$ [\%] & $S_{c}$ \\
\hline
L1 & 177 &  73.4 &	0.9 & 164 &	 76.2 & 0.8\\
L2 & 103 &	62.1 &	0.9 &  92 &	 82.6 & 0.7\\
L3 & 166 &	69.3 &	0.9 & 154 &	 74.0 & 0.8\\
L4 &  76 &	73.7 &	0.9 &  67 &	 71.6 & 0.7\\
L5 &  23 &	73.9 &	0.5 &  20 &	 90.0 & 0.7\\
L6 &  38 &	76.3 &	0.4 &  35 &	 62.9 & 0.8\\
R1 &  54 &	64.8 &	0.7 &  43 &	 60.5 & 0.8\\
R2 &  27 &	48.1 &	0.7 &  24 &	 75.0 & 0.4\\
R3 &  32 &  56.3 &  0.7 &  30 &  50.0 & 0.8\\
R4 &   8 &	62.5 &	0.4 &   7 &	 71.4 & 0.7\\
R5 &  17 &  52.9 &	0.7 &  14 &	 64.3 & 0.8\\
R6 &  38 &  73.7 &	0.7 &  32 &	 81.3 & 0.7\\
R7 &   8 &  62.5 &	0.7 &   8 & 100.0 & 0.3\\
P1 &  15 &	66.7 &  0.4 &  14 &	 78.6 & 0.6\\
P2 &   5 &	40.0 &	0.5 &   5 &	 40.0 & 0.6\\
P3 &   9 &	33.3 &  0.4 &   9 &  44.4 & 0.7\\
M1 &  21 &	66.7 &	0.5 &  19 &	 36.8 & 0.6\\
M2 &  18 &  55.6 &	0.2 &  16 &	 62.5 & 0.7\\
M3 &   7 &	42.9 &	0.6 &   6 &	 33.3 & 0.6\\
\hline
\end{tabular}
\label{tab:intra-stats}
\end{table}
%%%%%%%%%%%%%%%%%%%%%%%%%%%%%%%%%%%

%%%%%%%%%%%%%%%%%%%%%%%%%%%%%%%%%%%%%%%%%%%%
\subsection{Community detection}
\label{ssec:comm-det}
%%%%%%%%%%%%%%%%%%%%%%%%%%%%%%%%%%
In any organization different units play different roles, with some units serving more similar function than others.
Thus, it is expected that certain units should be more tightly connected, forming what is called in network analysis {\it clusters}.
To determine the structure and clustering of our networks, we employ the community detection techniques discussed in Section~\ref{ssec:sna-comm}.  
The goal is to determine whether the community partitions of the networks built by projecting egos' self-reported interactions onto units is reflective of the true organizational structure.

To account for all of the reported inter-unit interactions, we collapse the directed ties by summing the respective weights (i.e., weight from A to B with weight B to A) to create undirected networks.
We note that this process might result in certain interactions being accounted for twice which, for our purposes, indicates that a given interaction between the two units is stronger.
The undirected networks used to identify the community structure of each network are shown in Fig.~\ref{fig:networks-plots}(c) and Fig.~\ref{fig:networks-plots}(d), respectively.
We then use the Louvain community detection algorithm to partition the networks. 

For the collaboration network, we find two top-level clusters, one including exclusively units of type L, the other including units of type R, P, and M.
Interestingly, the units of type L are never clustered with units of a different type while, at the highest level, all of the remaining units form a single cluster.
This corresponds to the binary division of units as laboratory and non-laboratory programs.
The second-level clustering, depicted in Fig.~\ref{fig:networks-plots}(c), further separates out units of type R.
By third-level, two of the P units get pulled out from the second-level cluster to form a separate community while the third P unit and M units remain together.

The units in the L community of the collaboration network, shaded in red in Fig.~\ref{fig:networks-plots}(c), are among the largest in terms of total responses.
The percentages of intra-unit ties of these units are also among the highest (see Table~\ref{tab:intra-stats}) and are fairly consistent between the nodes with an average of $71.5~\%$ (${\rm st. dev.} = 5.1~\%$).
The second community, shaded in blue in Fig.~\ref{fig:networks-plots}(c), is made up exclusively of R type units, with an average percentage of intra-unit ties of $60.1~\%$ (${\rm st. dev.} = 8.5~\%$).
The final community, shaded in yellow in Fig.~\ref{fig:networks-plots}(c), is made up of the three P units and the three M units.
The units within this community are among the smallest in terms of ego responses and have the lowest average percentage of intra-unit ties at $50.9~\%$ (${\rm st. dev.} = 14.2~\%$).

The second-level communities of the collaboration network are reflective of the organization structure, as discussed in Section~\ref{ssec:survey}.
The perfect overlap in the communities detected within the network and the organizational unit functions confirms that the network generated by projecting ego networks on units provides a reliable approximation of the organization structure. 

In the following paragraphs, we discuss the community structure found in the advice network.
Through the Louvain community detection algorithm, 3 communities are found in the top-level partition of the advice network.
These three communities resemble the communities identified in the second-level partition of the collaboration network, however with more mixing between units of different types.
At the second-level of the advice network partitioning, shown in Fig.~\ref{fig:networks-plots}(d), the P and M units are separated out from L units and form a separate cluster, and the majority R community breaks into two clusters of size 5 (including one M unit) and size 3 (all R units). 
By the third-level, the  network falls apart into a number of single-unit communities.

The first community in the second-level partition of the advice network is made up exclusively of L units, as seen in the region shaded in red in Fig.~\ref{fig:networks-plots}(d).
The percentage of intra-unit ties of the nodes within this community is more varied compared to the L community in the collaboration network, with an average percentage of intra-unit ties of $76.2~\%$ (${\rm st. dev.} = 9.3~\%$).
The next community is a two node community, shaded in orange in Fig. \ref{fig:networks-plots}(d), that is grouped with the L community in the top-level partition; it includes P3 and M1, which have $44.4~\%$ and $36.8~\%$ of intra-unit ties respectively.
The third community consists of four units of type R and one M unit, shaded in blue in Fig.~\ref{fig:networks-plots}(d), with an average intra-unit percentage of $76.6~\%$ (${\rm st. dev.} = 15.2~\%$).
The fourth community is made up of the remaining three R units, shaded in indigo in Fig. \ref{fig:networks-plots}(d), and is grouped with the third community in the top-level partition.
The average percentage of intra-unit ties for this community is $60.6~\%$ (${\rm st. dev.} = 10.7~\%$).
The fifth and final community in the advice network is made up of two units of type P and a single M unit, shaded in yellow in Fig. \ref{fig:networks-plots}(d).
These three units have a highly varying percentage of intra-unit ties between them, with an average of $50.6~\%$ (${\rm st. dev.} = 24.4~\%$).

The communities in the advice network are also, in part, representative of the organizational structure. 
However, the partition is more fragmented and results in more mixing of different types of units.
Within each community, the percentage of intra-unit ties has a greater variation than that of the collaboration network.
Thus, the advice network does not reflect the organizational structure as well as the collaboration network.

%%%%%%%%%%%%%%%%%%%%%%%%%%%%%%%%%%%%%%%%%%%%%%%%%%%%%%%%%%%%%%
\subsection{Community assignment stability}
%%%%%%%%%%%%%%%%%%%%%%%%%%%%%%%%%%%%%%%%%%%%
Since the network of units is based on employees' responses, which are unequally represented in the overall sample.
It is thus important to validate the community structure we established with one detected from network data that is representative of the true distribution of employees across units.
To accomplish this, we conduct an experiment, as discussed in Section~\ref{ssec:exp}, where we perform a stratified sampling ($N = 1000$) over our datasets to mimic an idealized collection of survey responses that reflects the distribution of employees among units.
Such sampling ensures that the ego distribution is representative of the true distribution of employees in the organization.  

%%%%%%%%%%%%%%%%%%%%%%%%%%%%%%%%%%%
\begin{figure}[t]
\centering
\includegraphics[width=\linewidth]{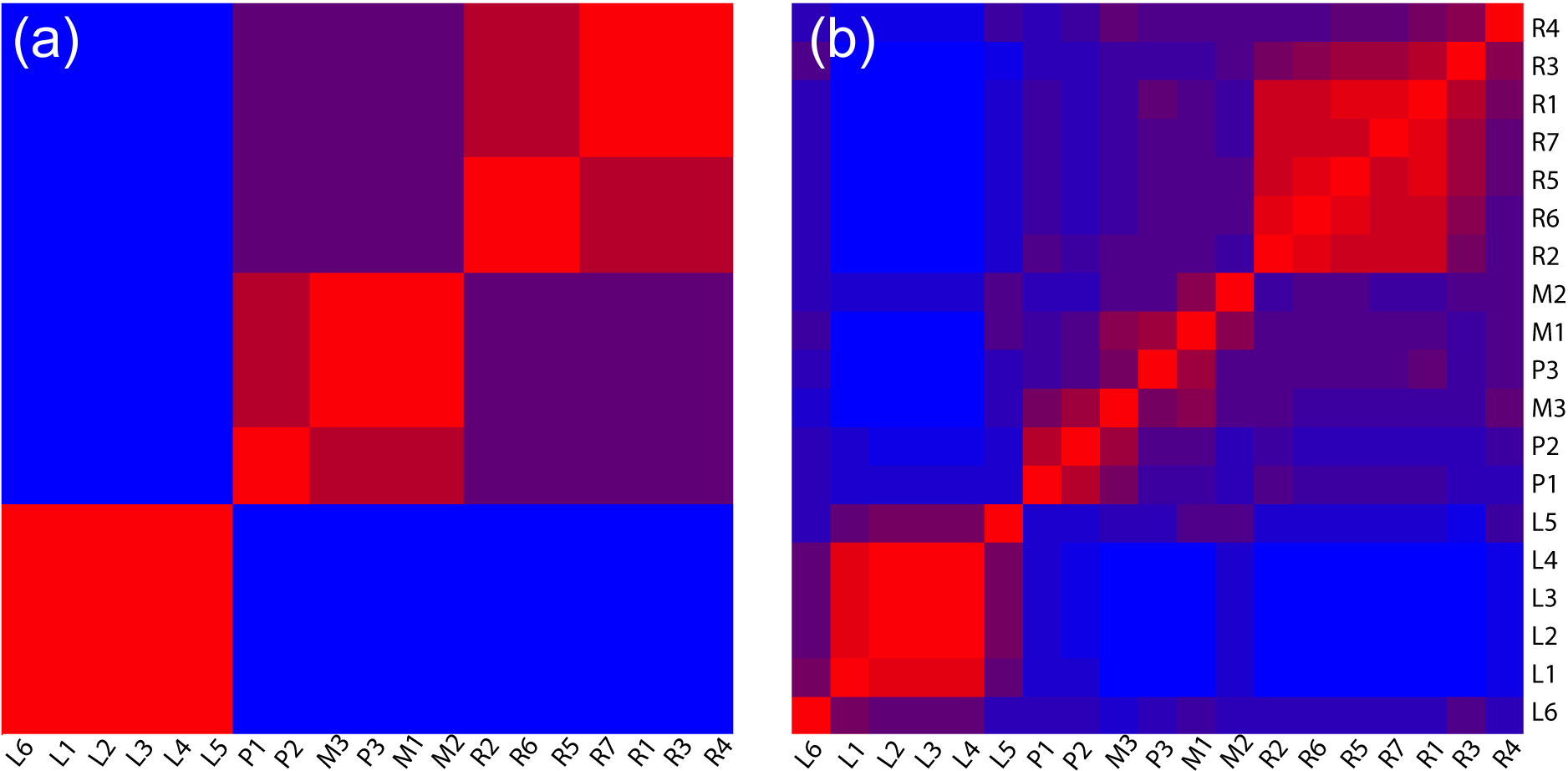}
\caption{A comparison of the hierarchical community structure between (a) the full collaboration network and (b) the cumulative community partitions of the re-sampled networks. 
The heatmap colors are placed on a range from blue (never in the same community) to red (completely nested in the same community). 
The structure identified from the Louvain algorithm applied to the full collaboration network was used to order the heatmap for the re-sampled networks.
The ordering of the matrix is determined by the hierarchical clustering algorithm applied to the collaboration network data.}
\label{fig:collab-heatmap}
\end{figure}
%%%%%%%%%%%%%%%%%%%%%%%%%%%%%%%%%%%

From each sample, a network of units is created.
The sampling method limits the overall sample size by the lowest responding unit, resulting in networks generated from $52~\%$ of the original data set.
Each network is then partitioned into communities based on the Louvain algorithm.

The community membership matrix for the collaboration network data can be is visualized in Figure \ref{fig:collab-heatmap}(b).
For comparison, the hierarchical partition of the organization network discussed in the previous section is displayed as a heatmap in Fig.~\ref{fig:collab-heatmap}(a).
While there are some communities that are consistent when re-sampling to obtain a representative sample, there are other communities that differ with the different distribution of egos.

As we discussed previously, the L community observed in the top-level partition is never broken down into sub-communities.
In the sampled networks, two of the L units -- L5 and L6 -- are not as frequently placed in the same community as the other labs.
In fact, L5 and L6 were only partitioned in communities with the other L units about half of the time.
This might in part result from the fact that these are the two smallest units within the L community, which leaves them more sensitive to sampling.
Still, these two L units tend to be clustered into the L community more frequently than into any other community.

To measure the community assignment stability, we use the percent difference between the average frequency of in- and out-of-community captured in the community membership matrix.
A value close to 1 indicates that a given unit is almost always assigned to the expected community while a value close to 0 indicates that the unit is just as frequently assigned to an unexpected community.
The assignment stability metric for each unit in the collaboration and advice networks can be seen in Table~\ref{tab:intra-stats}.
For the collaboration network, the units belonging to the L community have the highest assignment stability ($S_{c}=0.9$), with the exception of the two smaller units.
The core of the L community is stable under the sampling process.
The R community, as described in Section~\ref{ssec:comm-det}, is also fairly stable under the sampling process, with $S_{c}=0.7$ for all but one unit.
Finally, with the P+M community the assignment is quite less stable, with $S_{c}$ ranging from $0.6$ to $0.2$.

The assignment stability score tends to be smaller for smaller units, which may be an artifact of the sampling level used. 
Still, the two larger in terms of unit size communities are being reconstructed during the experiment, as can be seen in Fig.~\ref{fig:collab-heatmap}.
The purity of the partition of the sampled collaboration networks is $0.76$  (${\rm st.dev.}= 0.08$), and the F-measure is $0.72$ (${\rm st.dev.}= 0.09$), indicating similar network partitions to the community structure of the full network.
Thus, we expect that these two communities are representative of the true organization structure.

%%%%%%%%%%%%%%%%%%%%%%%%%%%%%%%%%%%
\begin{figure}[t]
\centering
\includegraphics[width=\linewidth]{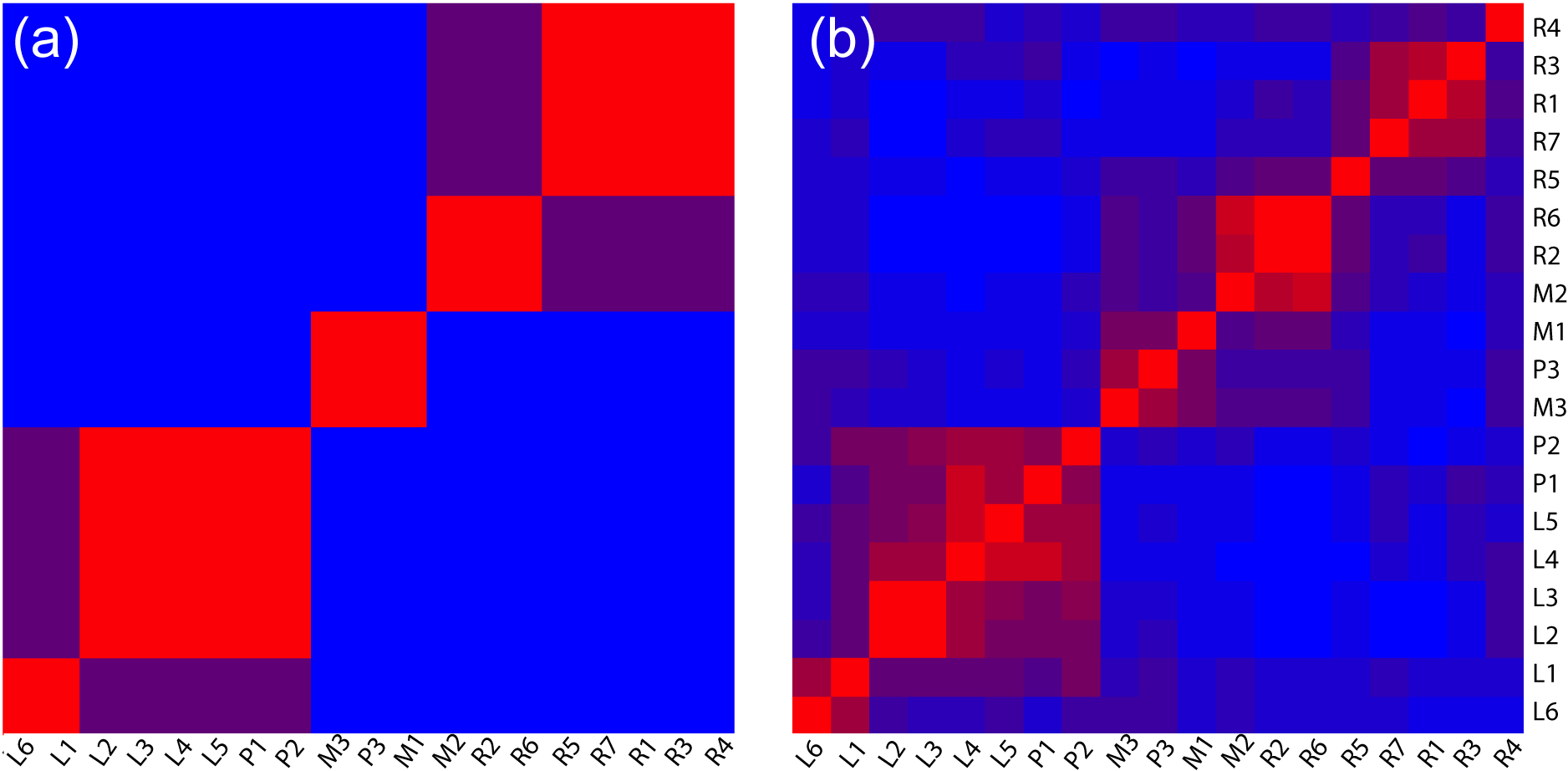}
\caption{A comparison of the hierarchical community structure between (a) the full advice network and (b) the cumulative community partitions of the re-sampled networks. 
The heatmap colors are placed on a range from blue (never in the same community) to red (completely nested in the same community). 
The structure identified from the Louvain algorithm applied to the full advice network was used to order the heatmap for the re-sampled networks.
The ordering of the matrix is determined by the hierarchical clustering algorithm applied to the advice network data.}
\label{fig:advice-heatmap}
\end{figure}
%%%%%%%%%%%%%%%%%%%%%%%%%%%%%%%%%%%

The same sampling process and analysis were performed on the advice networks dataset. 
The stratified sampling was again ran $N = 1000$ times.
The hierarchical community structure identified through the Louvain algorithm of the full data set and depicted in Fig.~\ref{fig:advice-heatmap}(a), however, is less present in the cumulative community sorting represented in the community membership matrix shown in Fig.~\ref{fig:advice-heatmap}(b).
The weaker community assignment stability is further confirmed by the stability score, with almost all units scoring between $S_{c}=0.6$ and $S_{c}=0.8$
While each unit's community assignment is somewhat stable, non of the communities detected in the advice network stands out as a highly stable, in contrast to the collaboration network.

The purity of the partition of the sampled advice networks is $0.71$  (${\rm st.dev.}= 0.09$), and the F-measure is $0.64$ (${\rm st.dev.}= 0.09$) -- similar to the purity and F-measure of the collaboration network sampling.
While the sampled network partitions are similar to the community structure of the full advice network, there are no individual communities that are highly stable.
On average, the community structure of the advice network may be reconstructed, each community individually are more susceptible to change during the sampling process.
Therefore, the exact communities identified in our dataset are less likely to be reproduced in a representative sample of egos.

%%%%%%%%%%%%%%%%%%%%%%%%%%%%%%%%%%%
\begin{figure*}[t]
\includegraphics{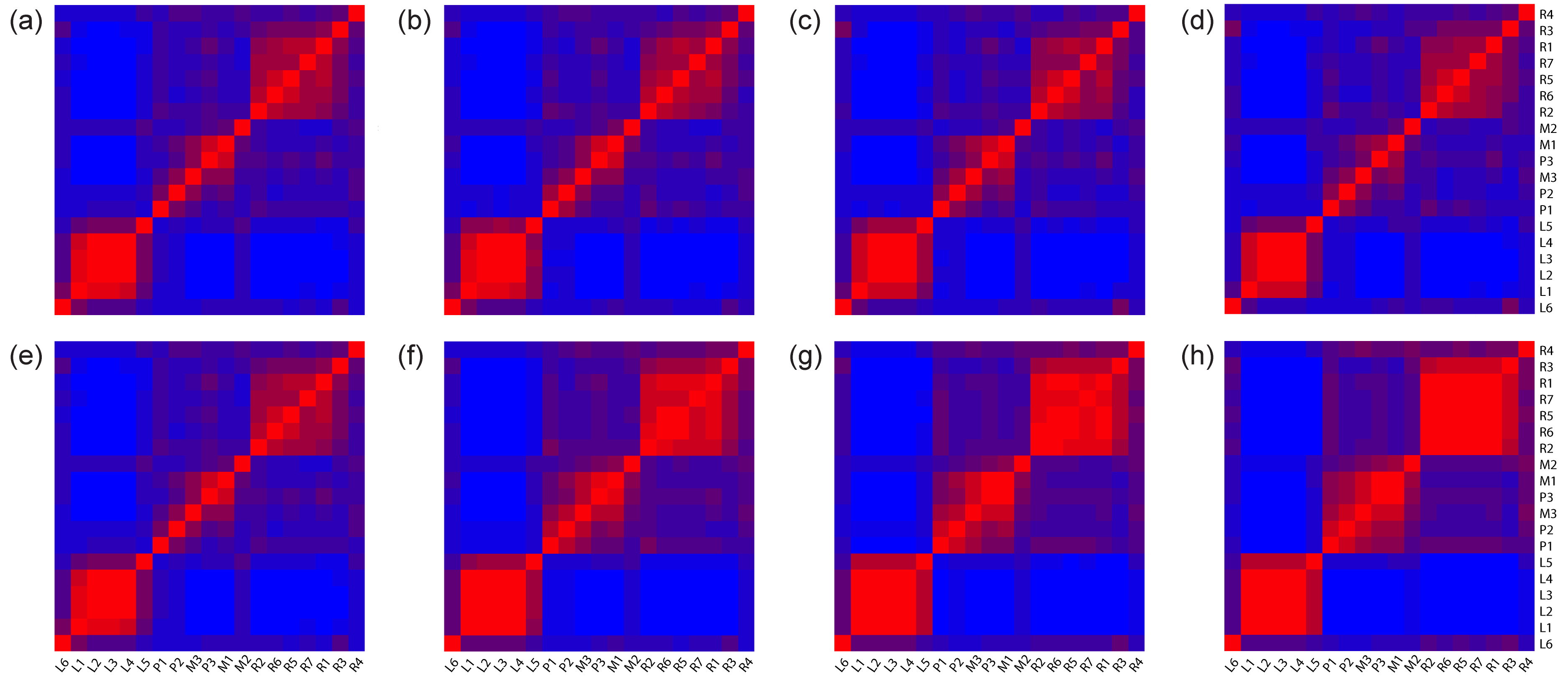}
\caption{Community Membership Matrices of the sampled networks for each time period. (a-d) The results of each time period sample's hierarchical community structure. (e-h) The cumulative results of each time period sample's hierarchical community structure.}
\label{fig:collab-exp}
\end{figure*}
%%%%%%%%%%%%%%%%%%%%%%%%%%%%%%%%%%%

The second investigation aims to to understand the effect of timed administration of our survey and cumulative data collection methods.
Surveys that obtain a lower response rate from certain units may result in disproportionately weak and fewer ties from those units within the organizational network.
To study this phenomena, we use the community assignment stability metric to quantify the effect of the underrepresentation of certain units.
We use the collaboration network for this analysis as it is the most representative of the true organizational structure.

In our analysis, we simulate quarterly survey collections in which the majority of units are sampled at $25~\%$ to imitate the response rate that our survey originally received.
In addition, at each stage, four randomly selected units are sampled at $15~\%$ to introduce the time-dependent differences in units' workflows.
The four resulting community membership matrices are shown in Fig.~\ref{fig:collab-exp}(a-d).

The L and R communities are mostly recovered in each of the sampled networks, with the average community assignment stability ranging from 0.6 to 0.8.
However, the P+M community is much less present in each of the community membership matrix visualizations, regardless of which units were under-sampled.
The average assignment stability of the P+M community is less than 0.5 in each of the survey periods, indicating that the P+M community is unstable at this sampling level.
The purity and F-measure range from $0.72$ to $0.74$ (each with ${\rm st.dev.}= 0.09$) and $0.68$ to $0.71$ (each with ${\rm st.dev.}= 0.10$) respectively, indicating that even at this low sampling rate the structure is somewhat recaptured.
While the whole partition of these sampled networks match relatively well to the full community structure, the individual communities differ in stability.

When the data is analyzed cumulatively -- that is by aggregating a given dataset with data from previous survey disseminations -- the communities become more stable over time, as can be seen in Fig.~\ref{fig:collab-exp}(e-h).
Whereas previously the P+M community was unstable, in the cumulative analysis, this community becomes clearly distinguishable and stable (average $S_{c} > 0.5$) by the second collection.
In each subsequent collection, the cumulative analysis results in an increase of both the average purity and F-measure of the partitions -- $0.74$, $0.79$, $0.81$, to $0.82$ (each with ${\rm st.dev.}= 0.09$).
This increase was also seen in the average F-measure -- $0.71$, $0.77$, $0.79$, to $0.81$ (each with ${\rm st.dev.}= 0.10$).
The extent to which the partition of the sampled partitions more closely match the full community structure increased with cumulative analysis.

At the response rates selected, certain features of the original unit partition are stable.
However, more than one survey administration period was needed in order to ensure the stability of all of the communities.
With the addition of the data collected in each survey period, the purity and F-measure of the full partition increased.
Through cumulative data collection and survey administration timing that capitalize on the workflow of an organization, the accurate organizational structure can be captured through modular analysis.

%%%%%%%%%%%%%%%%%%%%%%%%%%%%%%%%%%%%%%%%%%%%%%%%%%%%%%%%%%%%%%
\section{Discussion}\label{sec:discussion}
%%%%%%%%%%%%%%%%%%%%%%%%%%%%%%%%%%%%%%%%%%%%
Visualizing and analyzing formal and informal relationships within any organization can help shape business and operational strategy to maximize organic exchange of information, thereby helping the organization to become more sustainable and effective \citep{Gulati02-ONB, Cross04}.
However, ONA typically requires personally identifiable information in order to build the organizational network, which limits its utility and increases risk to participants \citep{Bell07-EMR}. 

The goal of this work is to explore the extent to which ego-centric data can be used to assess the interconnectedness within an organization.
Using survey data that captures two different types of interactions, we build projected networks of the organization's units.
We then perform a modular analysis of the resulting networks. 
As expected, the aspects of the organizational structure captured by the network built from interactions related to the organization's mission -- that is the collaboration network -- provides a reliable model of the true organizational structure.
The communities in the network resulting from the advice-related interaction, on the other hand, differ from that predefined structure, confirming that the aspects of the structure captured by the network depend on the type of examined interaction.
Thus, a particular care should be taken when designing the survey question about ego interactions to ensure that the target aspect of the organizational interconnectedness is properly assessed.

In order to confirm the reliability of our results, we conduct two additional analyses that test the stability of the identified structures.
The first experiment is used to correct for an unequal distribution of responses among units.
In order to validate the communities we found in both the collaboration and advice networks, we use stratified sampling to analyze the stability of these communities with respect to representative ego samples.
In the collaboration network, two out of three communities turn out to be stable under the re-sampling, while the third community is found to be unstable.
The advice network resulted in moderately stable community assignments, with each community being smaller and including a greater variety of units than those in the collaboration network.
This experiment validates the partition of our networks and gave insight into the communities that would be consistent given a representative sample of egos.

The second experiment was designed to determine how response rates that are dependent on timing of survey administration may impact the community structure of the resulting network.
Through analysis of the collaboration network, we found that most communities are stable even when a randomly selected fraction of units are under-sampled.
However, one community became stable only after cumulative analysis of two consecutive simulated survey administrations.
Thus, while the community structure of the organizational network is recoverable under low response rates, particular aspects of the structure are susceptible to an unequal distribution of the response rate.
This effect can be mitigated through the repeated collection of data, as in our experiment, or strategically stratified survey distribution that corresponds to the workflow of the organization.

%%%%%%%%%%%%%%%%%%%%%%%%%%%%%%%%%%%%%%%%%%%%%%%%%%%%%%%%%%%%%%
\subsection{Limitations}\label{ssec:limitations}
%%%%%%%%%%%%%%%%%%%%%%%%%%%%%%%%%%%%%%%%%%%%
The most important limitation of the proposed approach is that it does not allow to construct networks of individuals.
While the projected network correctly captures the organization's structure at the unit level, it does not consider the individual's connectedness within the organization.
As a result, analysis of certain organizational aspects, such as power structures \citep{Ramos19-AOP, Brass12, Krackhardt90-APL}, organizational commitment \citep{Lee11-ERC, Olfat20-OCI}, or the flow of information and knowledge between individuals \citep{Liebowitz05-LSN, Haythornthwaite96-SIE, Diez14-SCK}, are not possible. 

Moreover, while we anonomyze units prior to analysis, some analyses might require certain attributes of units, leaving them potentially identifiable to internal management.
This includes identifying units that play an important role in supporting network functions as well as units that function as a ``weak node", i.e., nodes that act as conduits  or bottlenecks in the organization \citep{Agneessens22-SCS}.
Thus, while the proposed approach mitigates risk to individuals within an organization, units and groups might still face potential harm.
The risk to identification is increased at each subsequent sub-level (e.g. large scale units compared to small working groups).

%%%%%%%%%%%%%%%%%%%%%%%%%%%%%%%%%%%%%%%%%%%%%%%%%%%%%%%%%%%%%%
\section{Conclusions}\label{sec:conclusions}
%%%%%%%%%%%%%%%%%%%%%%%%%%%%%%%%%%%%%%%%%%%%
To summarize, using our proposed approach, we have successfully created two networks of organizational units based on anonymous ego-centric data.
Importantly, this process required limited information about the egos and alters; in our case only the unit affiliation was needed to generate each network.
While this was not the focus of our study, additional ego and alter characteristics could have been incorporated into the analysis to answer more specific research questions.
For example, the gender of egos and alters could be used to investigate how the gendered interactions contribute to the formation of unit ties and communities, reflecting possible gendered differences in interpersonal networks \citep{Forret04-NBC, Burke95-INM}.

Additionally, other characteristics that were not collected as part of the NIST Interactions Survey could be collected to better understand different levels of connectedness -- such as division or group affiliation within units -- to reveal the more ephemeral interaction structures within units arising from particular coordinated tasks (the so-called short-term patterns) \citep{Quintane13-SLS}.
However, to ensure anonymity, the characteristics gathered should be balanced as to not compromise the PII by making respondents from underrepresented groups identifiable.
This anonymity is essential for many reasons, including protecting identity, ensuring honest and complete answers from respondents, as well as meeting requirements of access to the organization.
This is required for access to government agencies in which PII often cannot be collected without government official approval, and is also important for analysis of universities or national labs in which the focus is understanding departments or research groups rather than individuals.

The methodology proposed here sets a path forward for performing ONA that requires fairly accessible ego-centric data to investigate large-scale aspects of organizational structures.

%%%%%%%%%%%%%%%%%%%%%%%%%%%%%%%%%%%%%%%%%%%%%%%%%%%%%%%%%%%%%%
\section*{Acknowledgments}
We would like to thank faculty members for facilitating data collection.
The research was funded by the NSF under the Division of Physics Award 1344247.
RPD was in part supported by the National Science Foundation Graduate Research Fellowship under Grant No. DGE 1840340.
The views and conclusions contained in this paper are those of the authors and should not be interpreted as representing the official policies, either expressed or implied, of the National Science Foundation, or the U.S. Government. 
The U.S. Government is authorized to reproduce and distribute reprints for Government purposes notwithstanding any copyright notation herein. 
Any mention of commercial products is for information only; it does not imply recommendation or endorsement by NIST.

%%%%%%%%%%%%%%%%%%%%%%%%%%%%%%%%%%%%%%%%%%%%%%%%%%%%%%%%%%%%%%
%% References with bibTeX database:

%%%%%%%%%%%%%%%%%%%%%%%%%%%%%%%%%%%%%%%%%%%%%%%%%%%%%%%%%%%%%%

\begin{thebibliography}{54}
\expandafter\ifx\csname natexlab\endcsname\relax\def\natexlab#1{#1}\fi
\expandafter\ifx\csname url\endcsname\relax
  \def\url#1{\texttt{#1}}\fi
\expandafter\ifx\csname doi\endcsname\relax
  \def\doi#1{\texttt{#1}}\fi
\expandafter\ifx\csname urlprefix\endcsname\relax\def\urlprefix{URL: }\fi
\expandafter\ifx\csname doiprefix\endcsname\relax\def\doiprefix{DOI: }\fi

\bibitem[{Agneessens and Labianca(2022)}]{Agneessens22-SCS}
Agneessens, F., Labianca, G., 2022. Collecting survey-based social network
  information in work organizations. Soc. Netw. 68, 31--47.

\bibitem[{Bavelas(1950)}]{Bavelas50-CPG}
Bavelas, A., 1950. Communication patterns in task-oriented groups. J. Acoust.
  Soc. Am. 22~(6), 725--730.

\bibitem[{Bell and Bryman(2007)}]{Bell07-EMR}
Bell, E., Bryman, A., 2007. The ethics of management research: an exploratory
  content analysis. Br. J. Manag. 18~(1), 63--77.

\bibitem[{Blau(2017)}]{Blau17}
Blau, P.~M., 2017. Exchange and power in social life. Routledge.

\bibitem[{Blondel et~al.(2008)Blondel, Guillaume, Lambiotte, and
  Lefebvre}]{Blondel08-FUC}
Blondel, V.~D., Guillaume, J.-L., Lambiotte, R., Lefebvre, E., 2008. Fast
  unfolding of communities in large networks. J. Stat. Mech.: Theory. Exp.
  2008~(10), P10008.

\bibitem[{Borgatti and Foster(2003)}]{Borgatti03-NPO}
Borgatti, S.~P., Foster, P.~C., 2003. The network paradigm in organizational
  research: A review and typology. J. Manage. 29~(6), 991--1013.

\bibitem[{Borgatti and Molina(2003)}]{Borgatti03-ESI}
Borgatti, S.~P., Molina, J.~L., 2003. Ethical and strategic issues in
  organizational social network analysis. J. Appl. Behav. Sci. 39~(3),
  337--349.

\bibitem[{Borgatti and Molina(2005)}]{Borgatti05-TEG}
Borgatti, S.~P., Molina, J.-L., 2005. Toward ethical guidelines for network
  research in organizations. Soc. Netw. 27~(2), 107--117.

\bibitem[{Brandes et~al.(2009)Brandes, Lerner, and Snijders}]{Brandes09-NES}
Brandes, U., Lerner, J., Snijders, T., 2009. Networks evolving step by step:
  Statistical analysis of dyadic event data. In: 2009 International Conference
  on Advances in Social Network Analysis and Mining. IEEE, pp. 200--205.

\bibitem[{Brass et~al.(2004)Brass, Galaskiewicz, Greve, and Tsai}]{Brass04-SNO}
Brass, D.~J., Galaskiewicz, J., Greve, H.~R., Tsai, W., 2004. Taking stock of
  networks and organizations: A multilevel perspective. Acad. Manage. J.
  47~(6), 795--817.

\bibitem[{Brass and Krackhardt(2012)}]{Brass12}
Brass, D.~J., Krackhardt, D.~M., 2012. Power, politics, and social networks in
  organizations. In: G.~R.~Ferris, D. C.~T. (Ed.), Politics in organizations.
  Routledge, New York, NY, Ch.~12, pp. 389--410.
\url{https://doi.org/10.4324/9780203197424}
\newline\doiprefix\doi{10.4324/9780203197424}

\bibitem[{Breiger(1974)}]{Breiger74-DPG}
Breiger, R.~L., 1974. The duality of persons and groups. Soc. Forces 53~(2),
  181--190.

\bibitem[{Burke et~al.(1995)Burke, Rothstein, and Bristor}]{Burke95-INM}
Burke, R.~J., Rothstein, M.~G., Bristor, J.~M., 1995. Interpersonal networks of
  managerial and professional women and men: descriptive characteristics. Women
  Manag. Rev.

\bibitem[{Carpenter et~al.(2012)Carpenter, Li, and Jiang}]{Carpenter12-SNO}
Carpenter, M.~A., Li, M., Jiang, H., 2012. Social network research in
  organizational contexts: A systematic review of methodological issues and
  choices. J. Manage. 38~(4), 1328--1361.

\bibitem[{Colombo et~al.(2011)Colombo, Laursen, Magnusson, and
  Rossi-Lamastra}]{Colombo11-OIF}
Colombo, M.~G., Laursen, K., Magnusson, M., Rossi-Lamastra, C., 2011.
  Organizing inter-and intra-firm networks: what is the impact on innovation
  performance? Ind. Innov. 18~(6), 531--538.

\bibitem[{Cook and Emerson(1978)}]{Cook78-PEC}
Cook, K.~S., Emerson, R.~M., 1978. Power, equity and commitment in exchange
  networks. Am. Sociol. Rev., 721--739.

\bibitem[{Cronin et~al.(2021)Cronin, Perra, Rocha, Zhu, Pallotti, Gorgoni,
  Conaldi, and De~Vita}]{Cronin21-EIN}
Cronin, B., Perra, N., Rocha, L. E.~C., Zhu, Z., Pallotti, F., Gorgoni, S.,
  Conaldi, G., De~Vita, R., 2021. Ethical implications of network data in
  business and management settings. Soc. Netw. 67, 29--40.

\bibitem[{Cross et~al.(2004)Cross, Cross, and Parker}]{Cross04}
Cross, R.~L., Cross, R.~L., Parker, A., 2004. The hidden power of social
  networks: Understanding how work really gets done in organizations. Harvard
  Business Press.

\bibitem[{Csardi and Nepusz(2006)}]{igraph}
Csardi, G., Nepusz, T., 2006. The igraph software package for complex network
  research. InterJournal Complex Systems, 1695.

\bibitem[{Diener and Crandall(1978)}]{Diener78}
Diener, E., Crandall, R., 1978. Ethics in social and behavioral research. U
  Chicago Press, Oxford, England.

\bibitem[{D{\'\i}ez-Vial and Montoro-S{\'a}nchez(2014)}]{Diez14-SCK}
D{\'\i}ez-Vial, I., Montoro-S{\'a}nchez, {\'A}., 2014. Social capital as a
  driver of local knowledge exchange: A social network analysis. Knowl. Manag.
  Res. Pract. 12~(3), 276--288.

\bibitem[{Ellison and Langhout(2017)}]{Ellison17-SMR}
Ellison, E.~R., Langhout, R.~D., 2017. Sensitive topics, missing data, and
  refusal in social network studies: An ethical examination. Am. J. Community
  Psychol. 60~(3-4), 327--335.

\bibitem[{Espinal et~al.(2021)Espinal, Young, and Zwolak}]{Espinal21-NIR}
Espinal, L., Young, C., Zwolak, J.~P., 2021. Mapping employee networks through
  the nist interactions survey. {NIST Interagency/Internal Report (NISTIR)},
  National Institute of Standards and Technology, Gaithersburg, MD.
\newline\doiprefix\doi{10.6028/NIST.IR.8375}

\bibitem[{Flap et~al.(1998)Flap, Bulder, and V\"olker}]{Flap98-ION}
Flap, H., Bulder, B., V\"olker, B., Jun 1998. Intra-organizational networks and
  performance: A review. Comput. Math. Organ. Theory. 4~(2), 109--147.

\bibitem[{Forret and Dougherty(2004)}]{Forret04-NBC}
Forret, M.~L., Dougherty, T.~W., 2004. Networking behaviors and career
  outcomes: differences for men and women? J. Organ. Behav. 25~(3), 419--437.

\bibitem[{Ghawi and Pfeffer(2022)}]{Ghawi22-CMS}
Ghawi, R., Pfeffer, J., 2022. A community matching based approach to measuring
  layer similarity in multilayer networks. Soc. Netw. 68, 1--14.

\bibitem[{Gulati et~al.(2002)Gulati, Dialdin, and Wang}]{Gulati02-ONB}
Gulati, R., Dialdin, D.~A., Wang, L., 2002. Organizational networks. In: Baum,
  J. A.~C. (Ed.), The Blackwell Companion to Organizations. John Wiley \& Sons,
  Ltd, Oxford, England, Ch.~12, pp. 281--303.
\url{https://doi.org/10.1002/9781405164061.ch12}
\newline\doiprefix\doi{10.1002/9781405164061.ch12}

\bibitem[{Harris(2008)}]{Harris08-CAC}
Harris, J.~K., 2008. Consent and confidentiality: exploring ethical issues in
  public health social network research. Connections 28~(2), 81--96.

\bibitem[{Haythornthwaite(1996)}]{Haythornthwaite96-SIE}
Haythornthwaite, C., 1996. Social network analysis: An approach and technique
  for the study of information exchange. Libr. Inf. Sci. Res. 18~(4), 323--342.

\bibitem[{Kilduff and Brass(2010)}]{Kilduff10-OSN}
Kilduff, M., Brass, D.~J., 2010. Organizational social network research: Core
  ideas and key debates. Acad. Manag. Ann. 4~(1), 317--357.

\bibitem[{Krackhardt(1990)}]{Krackhardt90-APL}
Krackhardt, D., 1990. Assessing the political landscape: Structure, cognition,
  and power in organizations. Adm. Sci. Q. 35~(2), 342--369.

\bibitem[{Lazega et~al.(2012)Lazega, Mounier, Snijders, and
  Tubaro}]{Lazega12-NSD}
Lazega, E., Mounier, L., Snijders, T., Tubaro, P., 2012. Norms, status and the
  dynamics of advice networks: A case study. Soc. Netw. 34~(3), 323--332.

\bibitem[{Lazer and Katz(2003)}]{Lazer03-BEN}
Lazer, D., Katz, N., 2003. Building effective intra-organizational networks:
  The role of teams. In: Center for Public Leadership Working Paper Series.
  Center for Public Leadership, Ch.~3, pp. 83--107.
\url{http://hdl.handle.net/1721.1/55801}

\bibitem[{Lee and Kim(2011)}]{Lee11-ERC}
Lee, J., Kim, S., 2011. Exploring the role of social networks in affective
  organizational commitment: Network centrality, strength of ties, and
  structural holes. Am. Rev. Public Adm. 41~(2), 205--223.

\bibitem[{Liebowitz(2005)}]{Liebowitz05-LSN}
Liebowitz, J., 2005. Linking social network analysis with the analytic
  hierarchy process for knowledge mapping in organizations. J. Knowl. Manag.

\bibitem[{Merrill et~al.(2007)Merrill, Bakken, Rockoff, Gebbie, and
  Carley}]{Merrill07-ONA}
Merrill, J., Bakken, S., Rockoff, M., Gebbie, K., Carley, K.~M., 2007.
  Description of a method to support public health information management:
  Organizational network analysis. J. Biomed. Inform. 40~(4), 422--428.

\bibitem[{Moliterno and Mahony(2011)}]{Moliterno11-NTM}
Moliterno, T.~P., Mahony, D.~M., 2011. Network theory of organization: A
  multilevel approach. J. Manage. 37~(2), 443--467.

\bibitem[{Morris(2015)}]{Morris15}
Morris, M., 2015. Professional judgment and ethics. In: Scott, V., Wolfe, S.
  (Eds.), Community Psychology: Foundations for Practice. SAGE Publications,
  Ltd, 55 City Road, London, Ch.~5, pp. 132--156.
\url{https://sk.sagepub.com/books/community-psychology-foundations-for-practice}
\newline\doiprefix\doi{10.4135/9781483398150}

\bibitem[{Newman(2004)}]{Newman04-AWN}
Newman, M. E.~J., Nov 2004. Analysis of weighted networks. Phys. Rev. E 70,
  056131.

\bibitem[{Newman and Girvan(2004)}]{Newman04-FEC}
Newman, M. E.~J., Girvan, M., 2004. Finding and evaluating community structure
  in networks. Phys. Rev. E 69~(2), 026113.

\bibitem[{Nijhawan et~al.(2013)Nijhawan, Janodia, Muddukrishna, Bhat, Bairy,
  Udupa, and Musmade}]{Nijhawan13-IIC}
Nijhawan, L.~P., Janodia, M.~D., Muddukrishna, B.~S., Bhat, K.~M., Bairy,
  K.~L., Udupa, N., Musmade, P.~B., Jul 2013. Informed consent: Issues and
  challenges. J. Adv. Pharm. Technol. Res. 4~(3), 134--40.

\bibitem[{{NIST General Information}(2021)}]{NIST-info}
{NIST General Information}, 2021.
  \url{https://www.nist.gov/director/pao/nist-general-information}, online;
  accessed 17 November 2021.

\bibitem[{Olfat et~al.(2020)Olfat, Shokouhyar, Ahmadi, Tabarsa, and
  Sedaghat}]{Olfat20-OCI}
Olfat, M., Shokouhyar, S., Ahmadi, S., Tabarsa, G.~A., Sedaghat, A., Jan 2020.
  Organizational commitment and work-related implementation of enterprise
  social networks (esns): the mediating roles of employees' organizational
  concern and prosocial values. Online Inf. Rev. 44~(6), 1223--1243.

\bibitem[{Ozman(2009)}]{Ozman09-IFN}
Ozman, M., 2009. Inter-firm networks and innovation: a survey of literature.
  Econ. Innov. New Technol. 18~(1), 39--67.

\bibitem[{Provan et~al.(2007)Provan, Fish, and Sydow}]{Provan07-INL}
Provan, K.~G., Fish, A., Sydow, J., 2007. Interorganizational networks at the
  network level: A review of the empirical literature on whole networks. J.
  Manage. 33~(3), 479--516.

\bibitem[{Quintane et~al.(2013)Quintane, Pattison, Robins, and
  Mol}]{Quintane13-SLS}
Quintane, E., Pattison, P.~E., Robins, G.~L., Mol, J.~M., Oct 2013. Short- and
  long-term stability in organizational networks: Temporal structures of
  project teams. Soc. Netw. 35~(4), 528--540.

\bibitem[{{R Core Team}(2021)}]{R}
{R Core Team}, 2021. R: A Language and Environment for Statistical Computing. R
  Foundation for Statistical Computing, Vienna, Austria.
\url{https://www.R-project.org/}

\bibitem[{Raghavan et~al.(2007)Raghavan, Albert, and Kumara}]{Raghavan07-NLT}
Raghavan, U.~N., Albert, R., Kumara, S., 2007. Near linear time algorithm to
  detect community structures in large-scale networks. Phys. Rev. E 76~(3),
  036106.

\bibitem[{Ramos et~al.(2019)Ramos, Franco-Crespo, Gonz\'alez-P\'erez, Guerra,
  Ramos-Galarza, Pazmiño, and Tejera}]{Ramos19-AOP}
Ramos, V., Franco-Crespo, A., Gonz\'alez-P\'erez, L., Guerra, Y.,
  Ramos-Galarza, C., Pazmiño, P., Tejera, E., 2019. Analysis of organizational
  power networks through a holistic approach using consensus strategies.
  Heliyon 5~(2), e01172.

\bibitem[{Rosvall and Bergstrom(2008)}]{Rosvall08-MRW}
Rosvall, M., Bergstrom, C.~T., 2008. Maps of random walks on complex networks
  reveal community structure. Proc. Natl. Acad. Sci. U.S.A. 105~(4),
  1118--1123.

\bibitem[{Scott and Carrington(2011)}]{Scott11}
Scott, J., Carrington, P.~J., 2011. The SAGE handbook of social network
  analysis. SAGE publications.

\bibitem[{Tchalova and Eisenberger(2015)}]{Tchalova15-HBF}
Tchalova, K., Eisenberger, N., 2015. How the brain feels the hurt of
  heartbreak: Examining the neurobiological overlap between social and physical
  pain. In: Toga, A.~W. (Ed.), Brain Mapping. Academic Press, Waltham, pp.
  15--20.
\url{https://www.sciencedirect.com/science/article/pii/B9780123970251001445}
\newline\doiprefix\doi{https://doi.org/10.1016/B978-0-12-397025-1.00144-5}

\bibitem[{Warnes et~al.(2020)Warnes, Bolker, Bonebakker, Gentleman, Huber,
  Liaw, Lumley, Maechler, Magnusson, Moeller, Schwartz, and Venables}]{gplots}
Warnes, G.~R., Bolker, B., Bonebakker, L., Gentleman, R., Huber, W., Liaw, A.,
  Lumley, T., Maechler, M., Magnusson, A., Moeller, S., Schwartz, M., Venables,
  B., 2020. gplots: Various R Programming Tools for Plotting Data. R package
  version 3.0.4.
\url{https://CRAN.R-project.org/package=gplots}

\bibitem[{Wasserman and Faust(1994)}]{Wasserman94}
Wasserman, S., Faust, K., 1994. Social network analysis: Methods and
  applications. Cambridge university press.

\end{thebibliography}
\end{document}